# Chemical-Potential-Dependent Superconducting Spectroscopy on $Bi_2Te_3/NbSe_2$ Heterostructures

Pengyu Hu[1,†], Chun Yu Wan[2,†], Qiaoyan Yu[1,†], Chen Qian[3], Yisheng Gu[1], Xian Du[1], Xuhui Ning[1], Tingwen Miao[1], Shi Chen[1], Zihao Ma[1], Hao Yang[1,4,5], Hemian Yi[1,4,5], Weijiong Chen[1,4,5], Xiaoxue Liu[1,4,5], Liang Liu[1,4,5], Dandan Guan[1,4,5], Yaoyi Li[1,4,5], Shiyong Wang[1,4,5], Canhua Liu[1,4,5], Sangen Zhao[6], Fulin Wang[3], Junwei Liu[2], Hao Zheng[1,4,5,*] and Jinfeng Jia[1,4,5,6,7,*]

1. State Key Laboratory of Micro-nano Engineering Science, Tsung-Dao Lee Institute & School of Physics and Astronomy, Key Laboratory of Artificial Structures and Quantum Control (Ministry of Education), Shanghai Jiao Tong University, Shanghai 200240, China
2. Department of Physics, The Hong Kong University of Science and Technology, Hong Kong, China
3. National Engineering Research Center of Light Alloy Net Forming and State Key Laboratory of Metal Matrix Composite, School of Materials Science and Engineering, Shanghai Jiao Tong University, Shanghai 200240, China
4. Hefei National Laboratory, Hefei 230088, China
5. Shanghai Research Center for Quantum Sciences, 99 Xiupu Road, Shanghai 201315, China
6. Quantum Science Center of Guangdong-Hong Kong-Macao Greater Bay Area (Guangdong), Shenzhen 518045, China
7. Department of Physics, Southern University of Science and Technology, Shenzhen 518055, China

† These authors contributed equally to this work
* haozheng1@sjtu.edu.cn ; jfjia@sjtu.edu.cn

**Abstract:**
High-resolution scanning tunneling spectroscopy performed at dilution refrigerator temperatures was employed to investigate the superconducting properties of topological insulator/superconductor ($Bi_2Te_3/NbSe_2$) heterostructures. Coherence peaks with inner kinks or inner peaks, deviating from the conventional single s-wave spectral line shape, were observed in both 5 quintuple-layer (5QL) and 7QL $Bi_2Te_3$ thin films. Based on the combination of chemical-potential-dependent spectroscopic measurements and model simulations, we conclude that these spectral features arise from contributions of bulk and surface states in $Bi_2Te_3$ by selectively proximity-coupled to a single superconducting gap from $NbSe_2$, rather than inheriting the complicated superconductor's order parameter of $NbSe_2$.

**Main Text:**

The tunneling spectra on superconductors provide crucial insight into order parameter symmetries and quasiparticle excitations of the superconducting condensates. For example, conventional superconductors with a single isotropic s-wave gap display a characteristic "U"-shaped differential conductance (dI/dV) curve together with a single pair of coherence peaks. In contrast, the spectra of $NbSe_2$ show coherence peaks together with kink-like features at energies lower than the coherence peaks (the inner kinks), which have been interpreted as evidence for either a two-gap structure or a single anisotropic gap [1-10]. Unconventional superconductors such as $Bi_2Sr_2CaCu_2O_8$ and FeSe exhibit "V"-shaped tunneling spectra, a hallmark that has been linked to d-wave and s$\pm$-wave pairing symmetries, respectively [11-14]. Consequently, any observation of a tunneling spectrum that deviates from the single-pair coherence-peak, U-shaped gap form invariably draws research attention.

Topological insulator/superconductor (TI/SC) heterostructures have proven to be a practical platform for hosting novel physics, including the long-sought Majorana zero mode [15-29]. $Bi_2Te_3$ thin films of varying thicknesses have been grown on $NbSe_2$ substrates and characterized by scanning tunneling microscopy and spectroscopy (STM/STS) measurements at 300 mK. The resulting dI/dV spectra exhibit U-shaped gaps with a single pair of broad coherence peaks, with the gap size decreasing systematically as the TI film thickness increases [20]. Although the metallic bulk states help mediate the proximity effect, they may obscure contributions by the topological surface states. Despite the progress of thickness-dependent superconductivity, considerably less attention has been devoted to local chemical potential which dictates the contributions of metallic bulk states and topological surface states to the proximity-induced superconducting states.

Here, we employ a dilution-refrigerator-equipped STM system (base temperature of 40 mK) with enhanced energy resolution to revisit the superconducting spectra of $Bi_2Te_3/NbSe_2$ heterostructures. Fine structures within the superconducting coherence peaks, including inner kinks and inner peaks, can now be clearly resolved. The spectral shape and the relative weights of the coherence peaks and inner features exhibit a clear dependence on sample's chemical potential. Model simulations reveal that contributions from the proximity-induced superconductivity in both the bulk and surface states of $Bi_2Te_3$ account for the observed spectral shapes. This work uncovers a previously unexplored mechanism governing superconducting gap evolution and provides a new avenue for disentangling multiple superconducting channels in proximity-coupled quantum materials.

$Bi_2Te_3$ thin films were grown on $NbSe_2$ substrates at a substrate temperature of about 225 °C by molecular-beam epitaxy. During growth, the Bi and Te sources were heated to approximately 440 °C and 170 °C, respectively, using K-cell sources. The growth, transfer, and subsequent STM measurements were performed under ultrahigh-vacuum conditions, with a base pressure of $\sim 10^{-10}\ torr$. STM measurements were carried out at temperatures down to 40 mK using a dilution refrigerator. The STM tip was an Ag-coated tungsten tip, prepared by electrochemical etching of a tungsten wire.

$Bi_2Te_3$ is a prototypical TI material [30, 31], comprising a quintuple-layer (QL) stack with the Te-Bi-Te-Bi-Te sequence [Fig. 1(a)]. The thickness of 1 QL $Bi_2Te_3$ is about 1 nm. 2H-$NbSe_2$ is a SC compound made of Se-Nb-Se trilayers [Fig. 1(b)]. After vacuum cleaving, we obtain an atomically clean and flat $NbSe_2$ surface. dI/dV measurement in Fig. 1 (c) shows its SC spectrum with coherence peaks and inner kinks. We fit the tunnelling spectrum of $NbSe_2$ by the Dynes model [32] in which tunnelling conductance obeys

$$\frac{dI}{dV}=\frac{1}{2\pi}\int dE\int_0^{2\pi}d\theta\,\frac{df(E)}{dE}|_{E+eV}\text{abs}\left(\text{Re}\left[\frac{E-i\Gamma}{\sqrt{(E-i\Gamma)^2-\Delta(\theta)^2}}\right]\right)$$

with a gap function $\Delta(\theta)=\Delta_0+\Delta_1\cos(6\theta)$. The gap function corresponds to a single anisotropic gap, where $\Delta_0$ is the mean of the anisotropic gap, $2\Delta_1$ is the difference between gap maximum and gap minimum, and $\cos(6\theta)$ term gives a six-fold rotation symmetry. $f(E)$ is Fermi distribution function and $\Gamma$ is scattering rate. Through the fitting, we obtain that $\Delta_0=$ 1.03 meV, $\Delta_1=$ 0.256 meV, effective temperature $k_B T_{\text{eff}}=0.00345$ meV, $\Gamma=0.039$ meV. We then grew $Bi_2Te_3$ thin films on $NbSe_2$. The enhanced energy resolution of our dilution-refrigerator-equipped STM enables us to clearly resolve that the previously observed single broad coherence peak [33-35] actually consists of fine structures located at energies lower than the coherence peak, i.e. inner peaks, on our high-quality 5QL and 7QL $Bi_2Te_3$ samples, as shown in Fig. 1(d) and Fig. 1(e), respectively. At first glance, one may attribute the multiple-peak spectral shapes on the $Bi_2Te_3$ surface to the inheritance of $NbSe_2$'s anisotropic superconducting gap via the proximity effect. However, our subsequent research reveals an alternative interpretation.

Besides the spatially uniform, high-quality $Bi_2Te_3/NbSe_2$ samples demonstrated in Fig. 1, we have also prepared special 5QL $Bi_2Te_3$ thin films. Figure 2(a) shows such a sample containing two distinct rotation domains separated by an array of edge dislocations that form a domain boundary. The position-dependent dI/dV spectra, i.e. dI/dV(x,V) diagrams in Figs. 2(b) and 2(c) depict the spatial evolution of the TI gaps and SC gaps, respectively, in the two $Bi_2Te_3$ domains. Notably, the conduction-band minimum ($E_C$) is -60 meV in one domain [Fig. 2(d)] but -38meV in the other [Fig. 2(e)], proving that the chemical potentials ($\mu$) differ between domains. The SC spectral shapes also vary between domains. In the domain with high $\mu$, only a coherence peak located at -0.6 meV [marked by 1 in Fig. 2(f)] appears. However, in the low-$\mu$ domain, in addition to the coherence peak at -0.63 meV [marked by 1 in Fig. 2(g)], an inner kink at -0.35 meV [marked by 2 in Fig. 2(g)] also emerges, making the spectral shapes similar to that shown in Fig. 1(d).

To determine whether this phenomenon is specific to rotational domains, we performed analogous measurements on a special 7QL $Bi_2Te_3/NbSe_2$ sample. As shown in Fig. 3(a), no obvious domain structure is present in the STM image. However, the dI/dV(x,V) diagrams reveal spatial variations in both the wide-energy-scale electronic structure [Fig. 3(b)] and narrow-energy-scale superconducting spectra [Fig. 3(c)]. The single dI/dV spectra show that the local chemical potential varies at different locations, for example, $E_C$ = -80 meV and -29 meV in Figs. 3(d) and 3(e), respectively. Notably, the SC spectrum at the high-$\mu$ position features two peaks, with peaks 1 and 2 located at -0.65 meV and -0.27 meV, respectively [Fig. 3(f)]. The intensity of high energy peak (peak 1) is slightly stronger than that of inner peak (peak 2). In contrast, the SC spectrum at the low-$\mu$ location also features two peaks at the same energies, but peak 1 is now weaker than peak 2[Fig. 3(g)]. Our systematic investigation of the evolution of the SC spectral shapes confirms their dependence on the local chemical potential.

Motivated by the experimental observations, we now turn to a theoretical analysis. We adopt the minimal tight-binding model of TIs introduced in Ref. [16] and generalize it to a hexagonal lattice. Additional hopping terms are incorporated to enhance the hexagonal warping of the topological surface states in $Bi_2Te_3$ [36]. The normal-state Hamiltonian $H_{norm}(\boldsymbol{k})$ comprises four terms,

$$\begin{gathered} H_{norm}(\boldsymbol{k}) = \tau_x[d_1(\boldsymbol{k})s_1 + d_2(\boldsymbol{k})s_2 + d_3(\boldsymbol{k})s_3 + d_4(\boldsymbol{k})s_z] \\ +m(\boldsymbol{k})\tau_z s_0 + [a(\boldsymbol{k}) - \mu]\tau_0 s_0, \end{gathered} \tag{1}$$

with

$$d_1(\boldsymbol{k}) = 2t_1 \sin(k_y), \tag{2}$$

$$d_2(\boldsymbol{k}) = 2t_1 \sin\left(\frac{\sqrt{3}k_x - k_y}{2}\right), \tag{3}$$

$$d_3(\boldsymbol{k}) = 2t_1 \sin\left(\frac{-\sqrt{3}k_x - k_y}{2}\right), \tag{4}$$

$$d_4(\boldsymbol{k}) = 2t_{1z}\sin(k_z) + 2t_2\left[\sin(k_y) + \sin\left(\frac{\sqrt{3}k_x - k_y}{2}\right) + \sin\left(\frac{-\sqrt{3}k_x - k_y}{2}\right)\right], \tag{5}$$

$$s_1 = s_y, \tag{6}$$

$$s_2 = \frac{\sqrt{3}s_x - s_y}{2}, \tag{7}$$

$$s_3 = \frac{-\sqrt{3}s_x - s_y}{2}, \tag{8}$$

$$m(\boldsymbol{k}) = m_0 + m_1 f_1(k_x, k_y) + m_{1z}\cos(k_z), \tag{9}$$

$$a(\boldsymbol{k}) = a_1 f_1(k_x, k_y). \tag{10}$$

Pauli matrices $\tau_i$ and $\sigma_i$ represent the orbital and spin basis, respectively, and $\tau_0$ and $\sigma_0$ are the identity matrices. $f_1(k_x, k_y)$ describes the nearest-neighbor coupling on a two-dimensional hexagonal lattice as

$$f_1(k_x, k_y) = \frac{1}{2}\sum_{n=0}^{5} \exp\left[\mathrm{i}\sin\left(\frac{n\pi}{3}\right)k_x + \mathrm{i}\cos\left(\frac{n\pi}{3}\right)k_y\right]$$

We chose $m_0 = 1500, m_1 = -500, m_{1z} = -500, a_1 = -344, t_1 = 100, t_{1z} = 150, t_2 = 1000$.

The band structures were calculated in the slab model with $N_z$ layers and open boundary condition in the $z$ direction, by using $N_z = 25$ and 35 to model 5QL and 7QL $Bi_2Te_3$ respectively. The $E_C$ is $-728$ and $-730$ for $N_z = 25$ and $N_z = 35$ respectively. The corresponding band structures are presented in Figs. 4(a) and 5(a).

The Bogoliubov-de Gennes Hamiltonian for the slab model takes the form

$$H_{BdG}(k_x, k_y) = \begin{pmatrix} H_{norm}(k_x, k_y) & \mathrm{i}\Delta(z)\tau_0\sigma_y \\ -\mathrm{i}\Delta(z)\tau_0\sigma_y & H_{norm}(k_x, k_y) \end{pmatrix} \quad (11)$$

which operates in an $8N_z$-dimensional Hilbert space, accounting for two Nambu (particle–hole) pseudospins, two orbitals, two spin projections, and $N_z$ layers. The proximity-induced pair correlation is modelled as a decaying superconducting gap [18],

$$\Delta(z) = \Delta_0 \left[1 - \tanh\left(\frac{z}{L_d}\right)\right] \quad (12)$$

where $L_d$ is the decaying length and $z = 0,1, \dots, N_z - 1$. We set $L_d = 26$ and $L_d = 34$ for 5QL $Bi_2Te_3$ and 7QL $Bi_2Te_3$, as shown in Fig. 4(b) and Fig. 5(b) respectively. Since the surface states localize near the top layer, the effective superconducting gap is $\Delta(z = N_z - 1)$, while the effective superconducting gap for the bulk states is the average of $\Delta(z)$.

The surface density of states is calculated according to

$$\rho_{\mathrm{SDOS}}(E) = \sum\nolimits_{k_x,k_y} \left|\left\langle \psi_{n,k_x,k_y} \middle| z = N_z - 1 \right\rangle\right|^2 \delta\left(E - E_{n,k_x,k_y}\right) \quad (13)$$

$$\delta(E) = \frac{1}{\sqrt{2\pi\eta}} \exp\left(-\frac{E^2}{2\eta^2}\right) \quad (14)$$

where $E_{n,k_x,k_y}$ and $\psi_{n,k_x,k_y}$ denote the eigenvalues and eigenvectors of the Bogoliubov-de Gennes Hamiltonian. The surface density of states is obtained by projecting the states onto the top layer and broadening the spectral function with a Gaussian of width $\eta$.

Our systemic simulations of chemical potential-dependent SC spectra on 5QL and 7QL samples are presented in Figs. 4 and 5, respectively. When the chemical potential only crosses the surface state of $Bi_2Te_3$ ($\mu < E_C$), as shown in Figs. 4(c) and 5(c), a narrow U-shaped SC gap with a single pair of coherence peaks (denoted as peak 2) is observed. As the chemical potential rises and exceeds $E_C$, the bulk state begins to emerge at the Fermi level, leading to the appearance of additional peaks (marked as 1) at higher energies [Figs. 4(d) and 5(d)]. With a further increase in the chemical potential, both sets of features remain at the same energy positions, but the spectral weight of the inner feature (marked as 2) progressively decreases, evolving from inner peaks to inner kinks, as clearly shown in Figs. 4(e)-4(g) and 5(e)-5(g). Finally, when the chemical potential is sufficiently high ($\mu \gg E_C$) that the bulk state dominates at the Fermi level, only a single pair of coherence peaks (peak 1) with a wider SC gap appears in the SC spectra, as seen in Figs. 4(h) and 5(h). Notably, the simulated curves are in good agreement with our experimental data [Figs. 4(f), 4(g), 5(d), and 5(f)]. We therefore know that the outer peaks and inner peaks (or kinks) originate from the bulk and surface states of $Bi_2Te_3$, which are proximitized to the SC $Bi_2Te_3$.

The combined experimental and theoretical results establish a unified physical picture. Variations of the chemical potential continuously tune the relative contributions of bulk and surface states at the Fermi level [37, 38]. Since these states experience different proximity-induced pairing strengths, the local density of states can be expressed as a weighted superposition of bulk and surface superconducting components [39]. The observed spectral evolution therefore reflects a continuous transfer of superconducting spectral weight between bulk and surface channels. When one component dominates, the spectrum appears as a single superconducting gap. When both contributions become comparable, a distinct double-gap structure emerges naturally [40].

Moreover, since the superconductivity in $Bi_2Te_3$ is induced by the $NbSe_2$ substrate, which itself features an anisotropic gap described by the gap function $\Delta(\theta) = \Delta_0 + \Delta_1 \cos(6\theta)$, we repeat the above simulation, now incorporating the gap anisotropy into the superconducting proximity effect, as

$$\Delta(k_x, k_y, z) = \Delta_0^{TB}[1 - \tanh(z/L_{d0})] + \Delta_1^{TB} f(k_x, k_y)[1 - \tanh(z/L_{d1})] \quad (15)$$

$$f(k_x, k_y) = \frac{1}{3}\left[\cos(\pi k_y) + 2\cos(\pi k_y/2)\cos(\sqrt{3}\pi k_x/2)\right].$$

Because $\cos(6\theta)$ spans the range [−1, 1], whereas $f(k_x, k_y)$ spans [−1/2, 1], we choose $\Delta_0^{TB} = \Delta_0 - \Delta_1/3$ and $\Delta_1^{TB} = 4\Delta_1/3$ so that the range of $\Delta(k_x, k_y, z = 0)$ is preserved [41]. Hereafter, we drop the superscripts and simply use $\Delta_0$ to denote $\Delta_0^{TB}$ in the simulations based on tight-binding models. In this model, the $Bi_2Te_3$ thin film experiences spatially decaying superconducting gaps with distinct decay lengths, $L_{d0}$ for the $\Delta_0$ component and $L_{d1}$ for the $\Delta_1$ component, $\Delta_0$ and $\Delta_1$ are 1.47 and 0.53 respectively. In Fig. 6, we present the surface density of states calculated with two sets of parameters, i.e. $L_{d1} = L_{d0}$ and $L_{d1} = L_{d0}/2$. We find that incorporating both the $\Delta_0$ and $\Delta_1$ components into the simulation makes the gap size and the separation between peaks 1 and 2 too large, which then deviates considerably from the experimental data. Therefore, based on the comparison between the simulations and the experiments [Figs. 4-6], we conclude that although the $NbSe_2$ substrate possesses an anisotropic gap, the $Bi_2Te_3$ thin film inherits only a single isotropic gap, namely $\Delta_0$. A possible interpretation lies in the momentum matching between the substrate and the thin film in the proximity effect. $NbSe_2$ hosts multiple Fermi pockets around both the $\Gamma$ and $K$ points, giving rise to a complex anisotropic gap, whereas $Bi_2Te_3$ has only a single Fermi pocket around the $\Gamma$ point and therefore naturally inherits a single gap from $NbSe_2$.

In summary, spatially resolved spectroscopic measurements reveal an intrinsic double-gap feature in our proximitized $Bi_2Te_3/NbSe_2$ heterostructures. Theoretical calculations show that this behavior arises from a chemical-potential-dependent redistribution of spectral weight between bulk and surface superconducting states. Unlike thickness tuning, which globally modifies the superconducting coupling, local chemical-potential variations provide a continuous and spatially resolved means to control the interplay between bulk and surface superconductivity, thus offering a valuable tool for engineering topological superconducting phases and manipulating Majorana excitations in future quantum devices [42, 43]. These findings offer new insight into the microscopic nature of proximity-induced superconductivity and open new avenues for the controlled realization of topological superconducting states.

**Acknowledgment:**

H.Z. acknowledges the Ministry of Science and Technology of China (Grant No. 2025YFE0201200). We thank also NSFC(Grants No.12488101, No. 92365302, No. 22325203, No. 12474156, No. 12474121, No. 92577203, No. 12574146, No.12574174), the Ministry of Science and Technology of China (Grant No. 2024YFA1410100), the Science and Technology Commission of Shanghai Municipality (Grants No. 2019SHZDZX01, No.24LZ1401000), Cultivation Project of Shanghai Research Center for Quantum Sciences (Grant No. LZPY2024-04), Pujiang Talent Program (Grants No. 24PJA042 and No. 24PJA062) and Quantum Science and Technology-National Science and Technology Major Project (Grant No. 2021ZD0302500) for financial support.

**Reference:**

[1] H. F. Hess, R. B. Robinson, R. C. Dynes, J. M. Valles, and J. V. Waszczak, Scanning-Tunneling-Microscope Observation of the Abrikosov Flux Lattice and the Density of States near and inside a Fluxoid, Phys. Rev. Lett. 62, 214 (1989).

[2] H. F. Hess, R. B. Robinson, and J. V. Waszczak, Vortex-Core Structure Observed with a Scanning Tunneling Microscope, Phys. Rev. Lett. 64, 2711 (1990).

[3] Z. Dai, Q. Xue, Y. Gong, C. G. Slough, and R. V. Coleman, Scanning-Probe-Microscopy Studies of Superlattice Structures and Density-Wave Structures in 2H-$NbSe_2$, 2H-$TaSe_2$, and 2H-$TaS_2$ Induced by Fe Doping, Phys. Rev. B 48, 14543 (1993).

[4] J. G. Rodrigo and S. Vieira, STM Study of Multiband Superconductivity in $NbSe_2$ Using a Superconducting Tip, Physica C: Superconductivity 404, 306 (2004).

[5] A. Kohen, T. Proslier, T. Cren, Y. Noat, W. Sacks, H. Berger, and D. Roditchev, Probing the Superfluid Velocity with a Superconducting Tip: The Doppler Shift Effect, Phys. Rev. Lett. 97, 027001 (2006).

[6] I. Guillamon, H. Suderow, F. Guinea, and S. Vieira, Intrinsic Atomic-Scale Modulations of the Superconducting Gap of 2H-$NbSe_2$, Phys. Rev. B 77, 134505 (2008).

[7] J. G. Rodrigo, V. Crespo, and S. Vieira, Scanning Tunneling Spectroscopy of the Vortex State in $NbSe_2$ Using a Superconducting Tip, Physica C 468, 547 (2008).

[8] Y. Noat, J. A. Silva-Guillén, T. Cren, V. Cherkez, C. Brun, S. Pons, F. Debontridder, D. Roditchev, W. Sacks, L. Cario, P. Ordejón, A. García, and E. Canadell, Quasiparticle Spectra of 2H-$NbSe_2$: Two-Band Superconductivity and the Role of Tunneling Selectivity, Phys. Rev. B 92, 134510 (2015).

[9] A. Sanna, C. Pellegrini, E. Liebhaber, K. Rossnagel, K. J. Franke, and E. K. U. Gross, Real-Space Anisotropy of the Superconducting Gap in the Charge-Density Wave Material 2H-$NbSe_2$, npj Quantum Mater. 7, 6 (2022).

[10] Z. Zhu, W. Cheng, D. Liu, P. Hu, Y. Yang, Q. Yu, S. Xue, R. Xi, X. Chen, J. Sun, D. Guan, Y. Li, S. Wang, C. Liu, Z. Xu, X. Liu, H. Zheng, and J. Jia, Supercurrent Effect in a Charge Density Wave Intertwined Superconductor, Nat. Commun. 17, 7009 (2026).

[11] K. Ichimura, K. Nomura, F. Minami, and S. Takekawa, STM Study of the Electronic Density of States in the Superconducting Phase of $Bi_2Sr_2CaCu_2O_8$, Solid State Commun. 82, 171 (1992).

[12] T. Kurosawa, T. Yoneyama, Y. Takano, M. Hagiwara, R. Inoue, N. Hagiwara, K. Kurusu,

K. Takeyama, N. Momono, M. Oda, and M. Ido, Large Pseudogap and Nodal Superconducting Gap in $Bi_2Sr_{2-x}La_xCuO_{6+\delta}$ and $Bi_2Sr_2CaCu_2O_{8+\delta}$: Scanning Tunneling Microscopy and Spectroscopy, Phys. Rev. B 81, 094519 (2010).
[13] L. Jiao, C.-L. Huang, S. Rößler, C. Koz, U. K. Rößler, U. Schwarz, and S. Wirth, Superconducting Gap Structure of FeSe, Sci. Rep. 7, 44024 (2017).
[14] Q. Gu, S. Wan, Q. Tang, Z. Du, H. Yang, Q.-H. Wang, R. Zhong, J. Wen, G. D. Gu, and H.-H. Wen, Directly Visualizing the Sign Change of D-Wave Superconducting Gap in $Bi_2Sr_2CaCu_2O_{8+\delta}$ by Phase-Referenced Quasiparticle Interference, Nat. Commun. 10, 1603 (2019).
[15] L. Fu and C. L. Kane, Superconducting Proximity Effect and Majorana Fermions at the Surface of a Topological Insulator, Phys. Rev. Lett. 100, 096407 (2008).
[16] P. Hosur, P. Ghaemi, R. S. K. Mong, and A. Vishwanath, Majorana Modes at the Ends of Superconductor Vortices in Doped Topological Insulators, Phys. Rev. Lett. 107, 097001 (2011).
[17] M.-X. Wang, C. Liu, J.-P. Xu, F. Yang, L. Miao, M.-Y. Yao, C. L. Gao, C. Shen, X. Ma, X. Chen, Z.-A. Xu, Y. Liu, S.-C. Zhang, D. Qian, J.-F. Jia, and Q.-K. Xue, Coexistence of superconductivity and topological order in $Bi_2Se_3$ thin films, Science 336, 52 (2012).
[18] Z.-Z. Li, F.-C. Zhang, and Q.-H. Wang, Majorana modes in a topological insulator/s-wave superconductor heterostructure, Sci. Rep. 4, 6363 (2014).
[19] S.-Y. Xu, N. Alidoust, I. Belopolski, A. Richardella, C. Liu, M. Neupane, G. Bian, S.-H. Huang, R. Sankar, C. Fang, B. Dellabetta, W. Dai, Q. Li, M. J. Gilbert, F. Chou, N. Samarth, and M. Z. Hasan, Momentum-space imaging of Cooper pairing in a half-Dirac-gas topological superconductor, Nat. Phys. 10, 943 (2014).
[20] J.-P. Xu, C. Liu, M.-X. Wang, J. Ge, Z.-L. Liu, X. Yang, Y. Chen, Y. Liu, Z.-A. Xu, C.-L. Gao, D. Qian, F.-C. Zhang, and J.-F. Jia, Artificial Topological Superconductor by the Proximity Effect, Phys. Rev. Lett. 112, 217001 (2014).
[21] Q. L. He, H. Liu, M. He, Y. H. Lai, H. He, G. Wang, K. T. Law, R. Lortz, J. Wang, and I. K. Sou, Two-Dimensional Superconductivity at the Interface of a $Bi_2Te_3$/FeTe Heterostructure, Nat. Commun. 5, 4247 (2014).
[22] H.-H. Sun, K.-W. Zhang, L.-H. Hu, C. Li, G.-Y. Wang, H.-Y. Ma, Z.-A. Xu, C.-L. Gao, D.-D. Guan, Y.-Y. Li, C. Liu, D. Qian, Y. Zhou, L. Fu, S.-C. Li, F.-C. Zhang, and J.-F. Jia, Majorana zero mode detected with spin selective Andreev reflection in the vortex of a topological superconductor, Phys. Rev. Lett. 116, 257003 (2016).
[23] J. A. Hlevyack, S. Najafzadeh, M.-K. Lin, T. Hashimoto, T. Nagashima, A. Tsuzuki, A. Fukushima, C. Bareille, Y. Bai, P. Chen, R.-Y. Liu, Y. Li, D. Flötotto, J. Avila, J. N. Eckstein, S. Shin, K. Okazaki, and T. C. Chiang, Massive suppression of proximity pairing in topological $(Bi_{1-x}Sb_x)_2Te_3$ films on niobium, Phys. Rev. Lett. 124, 236402 (2020).
[24] S. Wan, Q. Gu, H. Li, H. Yang, J. Schneeloch, R. D. Zhong, G. D. Gu, and H.-H. Wen, Twofold symmetry of proximity-induced superconductivity in $Bi_2Te_3/Bi_2Sr_2CaCu_2O_{8+\delta}$ heterostructures revealed by scanning tunneling microscopy, Phys. Rev. B 101, 220503 (2020).
[25] Y. Zang, F. Küster, J. Zhang, D. Liu, B. Pal, H. Deniz, P. Sessi, M. J. Gilbert, and S. S. P. Parkin, Competing energy scales in topological superconducting heterostructures, Nano Lett. 21, 2758 (2021).
[26] Z. Zhu, M. Papaj, X.-A. Nie, H.-K. Xu, Y.-S. Gu, X. Yang, D. Guan, S. Wang, Y. Li, C. Liu, J. Luo, Z.-A. Xu, H. Zheng, L. Fu, and J.-F. Jia, Discovery of segmented Fermi surface

induced by Cooper pair momentum, Science 374, 1381 (2021).
[27] H. Yi, L.-H. Hu, Y. Wang, R. Xiao, J. Cai, D. R. Hickey, C. Dong, Y.-F. Zhao, L.-J. Zhou, R. Zhang, A. R. Richardella, N. Alem, J. A. Robinson, M. H. W. Chan, X. Xu, N. Samarth, C.-X. Liu, and C.-Z. Chang, Crossover from Ising- to Rashba-type superconductivity in epitaxial $Bi_2Se_3$/monolayer $NbSe_2$ heterostructures, Nat. Mater. 21, 1366 (2022).
[28] K. Chen, M. Chen, C. Wen, Z. Hou, H. Yang, and H.-H. Wen, Superconductivity and vortex structure in $Bi_2Te_3/FeTe_{0.55}Se_{0.45}$ heterostructures with varying $Bi_2Te_3$ thickness, Phys. Rev. B 108, 184512 (2023).
[29] J. A. Hlevyack, S.-Y. Guan, Y.-C. Tsui, C.-A. Wang, W.-Y. Chen, M.-K. Lin, Y. Li, S. S. Babu, P. Chen, R.-Y. Liu, J.-E. Lee, J. Hwang, S. Tang, Y. Bai, D. Flötotto, S.-K. Mo, J. N. Eckstein, P. Abbamonte, T.-M. Chuang, and T.-C. Chiang, Topological surface superconductivity via Josephson coupling in $Bi_2Te_3$/Nb, Nano Lett. 25, 16387 (2025).
[30] Y. L. Chen, J. G. Analytis, J. H. Chu, Z. K. Liu, S. K. Mo, X. L. Qi, H. J. Zhang, D. H. Lu, X. Dai, Z. Fang, S. C. Zhang, I. R. Fisher, Z. Hussain, and Z. X. Shen, Experimental Realization of a Three-Dimensional Topological Insulator, $Bi_2Te_3$, Science 325, 178 (2009).
[31] D. Hsieh, Y. Xia, D. Qian, L. Wray, F. Meier, J. H. Dil, J. Osterwalder, L. Patthey, A. V. Fedorov, H. Lin, A. Bansil, D. Grauer, Y. S. Hor, R. J. Cava, and M. Z. Hasan, Observation of Time-Reversal-Protected Single-Dirac-Cone Topological-Insulator States in $Bi_2Te_3$ and $Sb_2Te_3$, Phys. Rev. Lett. 103, 146401 (2009).
[32] R. C. Dynes, V. Narayanamurti, and J. P. Garno, Direct Measurement of Quasiparticle-Lifetime Broadening in a Strong-Coupled Superconductor, Phys. Rev. Lett. 41, 1509 (1978).
[33] D. Liu, P.-Y. Cai, Y. Gu, Q. Yu, S. Xue, R. Xi, X. Chen, J. Sun, X. Du, Y. Zhang, X. Ning, T. Miao, P. Hu, H. Yang, D. Guan, X. Liu, L. Liu, Y. Li, S. Wang, C. Liu, Y. Zhou, H. Zheng, and J. Jia, Potential Fu-Kane topological superconductor island in $Bi_2Te_3/VSi_x$/Si nanostructure, Chin. Phys. Lett. 42, 037403 (2025).
[34] H. Zheng and J.-F. Jia, Topological Superconductivity in a $Bi_2Te_3/NbSe_2$ Heterostructure: A Review*, Chin. Phys. B 28, 067403 (2019).
[35] J.-P. Xu, M.-X. Wang, Z. L. Liu, J.-F. Ge, X. Yang, C. Liu, Z. A. Xu, D. Guan, C. L. Gao, D. Qian, Y. Liu, Q.-H. Wang, F.-C. Zhang, Q.-K. Xue, and J.-F. Jia, Experimental Detection of a Majorana Mode in the Core of a Magnetic Vortex inside a Topological Insulator-Superconductor $Bi_2Te_3/NbSe_2$ heterostructure, Phys. Rev. Lett. 114, 017001 (2015).
[36] L. Fu, Hexagonal Warping Effects in the Surface States of the Topological Insulator $Bi_2Te_3$. Phys. Rev. Lett. 103, 266801 (2009).
[37] K. Park, G. Csire, and B. Ujfalussy, Proximity Effect in a Superconductor-Topological Insulator Heterostructure Based on First Principles, Phys. Rev. B 102, 134504 (2020).
[38] P. Sitthison and T. D. Stanescu, Robustness of Topological Superconductivity in Proximity Coupled Topological Insulator Nanoribbons, Phys. Rev. B 90, 035313 (2014).
[39] W. Dai, A. Richardella, R. Du, W. Zhao, X. Liu, C. X. Liu, S.-H. Huang, R. Sankar, F. Chou, N. Samarth, and Q. Li, Proximity-Effect-Induced Superconducting Gap in Topological Surface States-A Point Contact Spectroscopy Study of $NbSe_2/Bi_2Se_3$ Superconductor-Topological Insulator Heterostructures, Sci. Rep. 7, 7631 (2017).
[40] A. Banerjee, A. Sundaresh, R. Ganesan, and P. S. A. Kumar, Signatures of Topological Superconductivity in Bulk-Insulating Topological Insulator $BiSbTe_{1.25}Se_{1.75}$ in Proximity with Superconducting $NbSe_2$, ACS Nano 12, 12665 (2018).

[41] J. A. Galvis, E. Herrera, C. Berthod, S. Vieira, I. Guillamón, and H. Suderow, Tilted vortex cores and superconducting gap anisotropy in 2H-$NbSe_2$, Commun. Phys. 1, 30 (2018).
[42] C.-K. Chiu, W. S. Cole, and S. Das Sarma, Induced Spectral Gap and Pairing Correlations from Superconducting Proximity Effect, Phys. Rev. B 94, 125304 (2016).
[43] T. D. Stanescu, J. D. Sau, R. M. Lutchyn, and S. Das Sarma, Proximity Effect at the Superconductor-Topological Insulator Interface, Phys. Rev. B 81, 241310 (2010).

**Figures:**

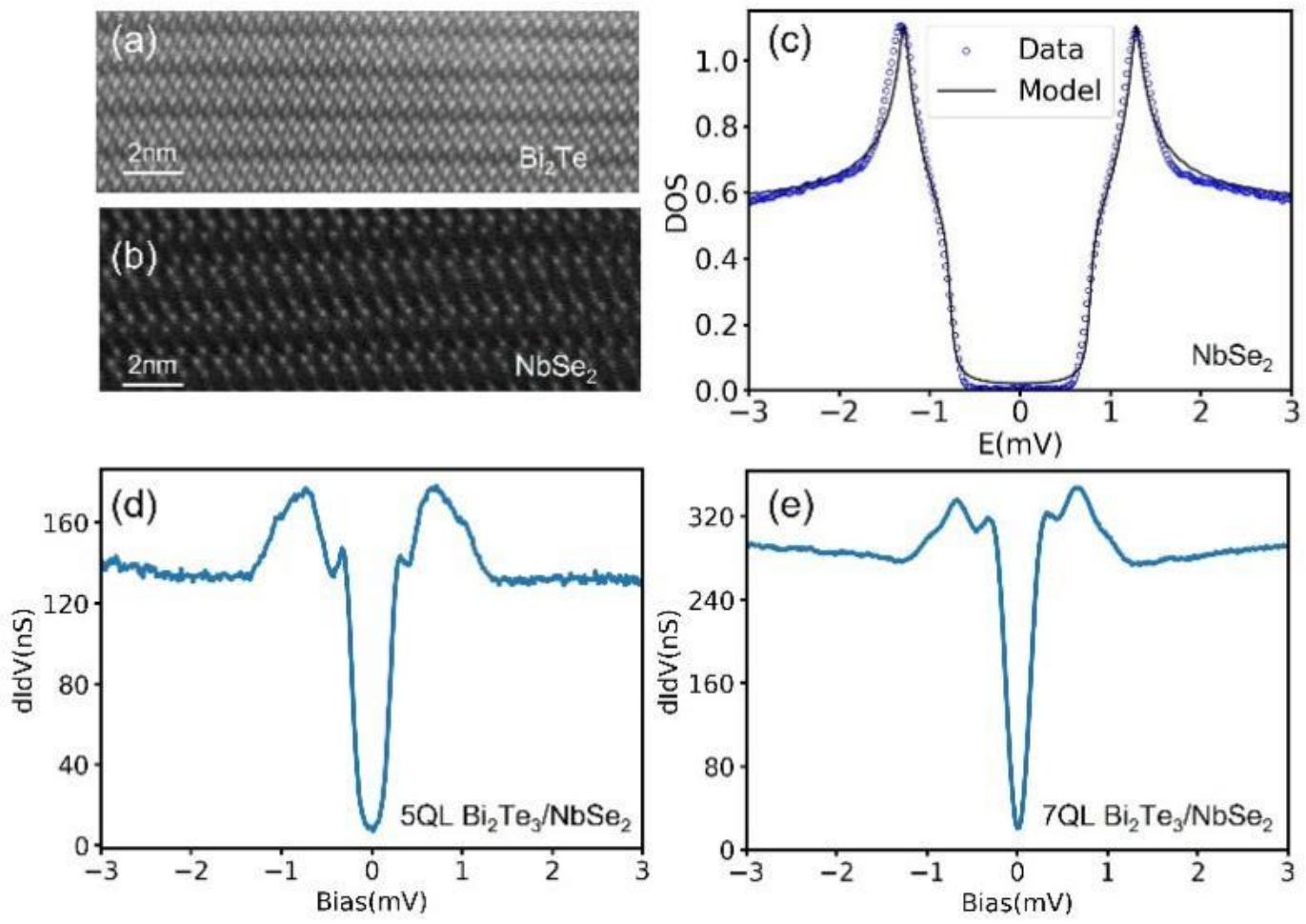

FIG. 1. (a) and (b) the high-angle annular dark-field scanning transmission electron microscopy (HAADF-STEM) images of $Bi_2Te_3$ thin film and $NbSe_2$ substrate, respectively, acquired using a field-emission aberration-corrected electron microscope. (c) dI/dV spectrum on $NbSe_2$. The dashed curve represents the Dynes-model fit to the superconducting gap structure. We obtain that $\Delta_0 = 1.03\ meV$, $\Delta_1 = 0.256\ meV$. (d) and (e) dI/dV spectrum on 5QL and 7QL $Bi_2Te_3$ film on $NbSe_2$ respectively.

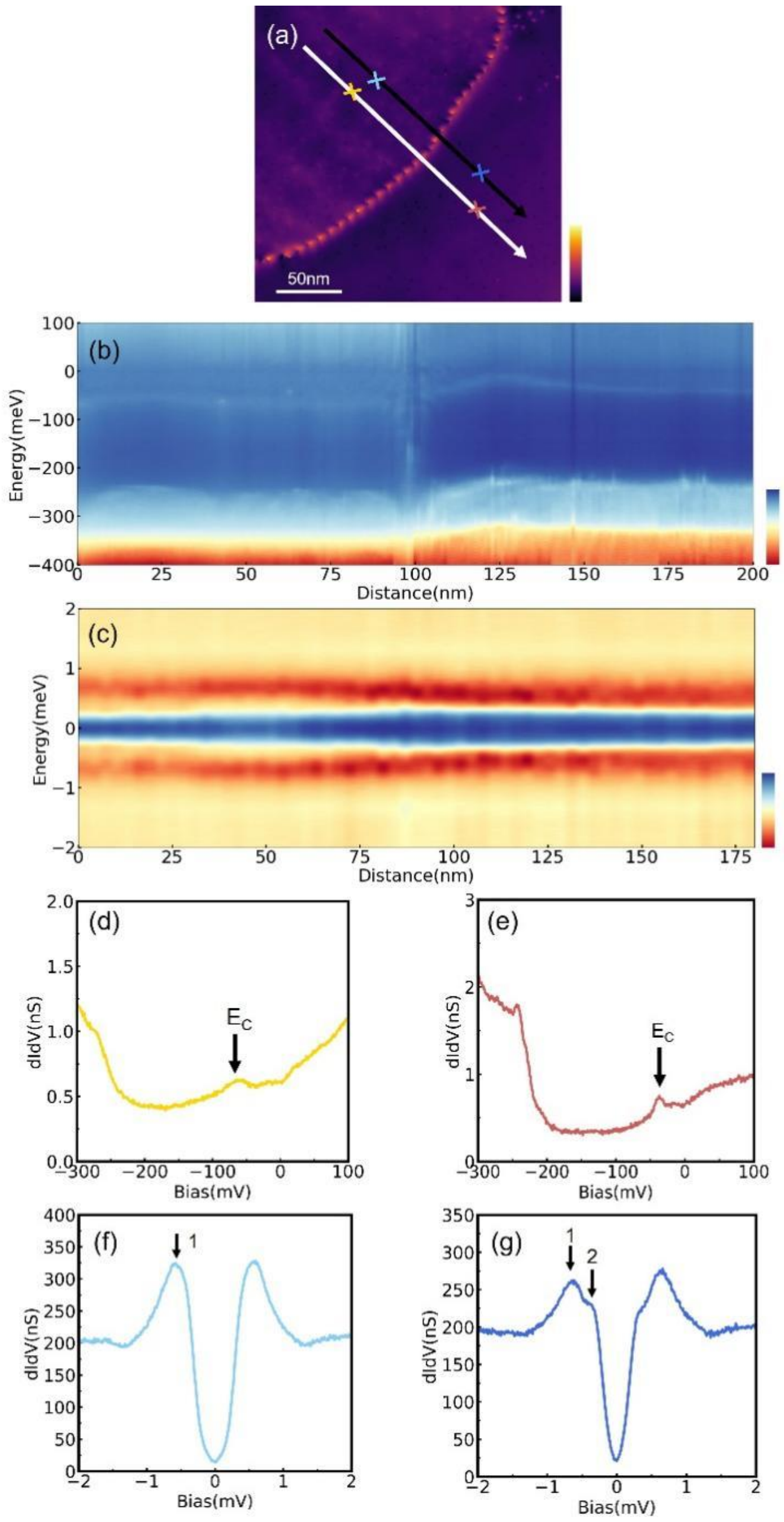

FIG. 2. (a) STM image of the 5QL $Bi_2Te_3$ containing two rotation domains. (b) and (c) spatially resolved dI/dV(x,V) diagrams acquired over a wide and narrow energy range along the white and black arrows in (a) respectively. (d) and (e) Representative dI/dV spectra taken on opposite sides of the domain boundary from the line cut in (b). The corresponding locations are marked by colored crosses in (a). The conduction-band minimum ($E_C$) marked by arrow is 60meV and 38meV below the Fermi Level on the left (d) and right (e) domain respectively. (f) and (g) Representative superconducting spectra extracted from the line cut in (c), illustrating the distinct spectral line shapes on the different domains. The measurement locations are indicated by colored crosses in (a). 1 and 2 indicate the coherent peak and inner kink.

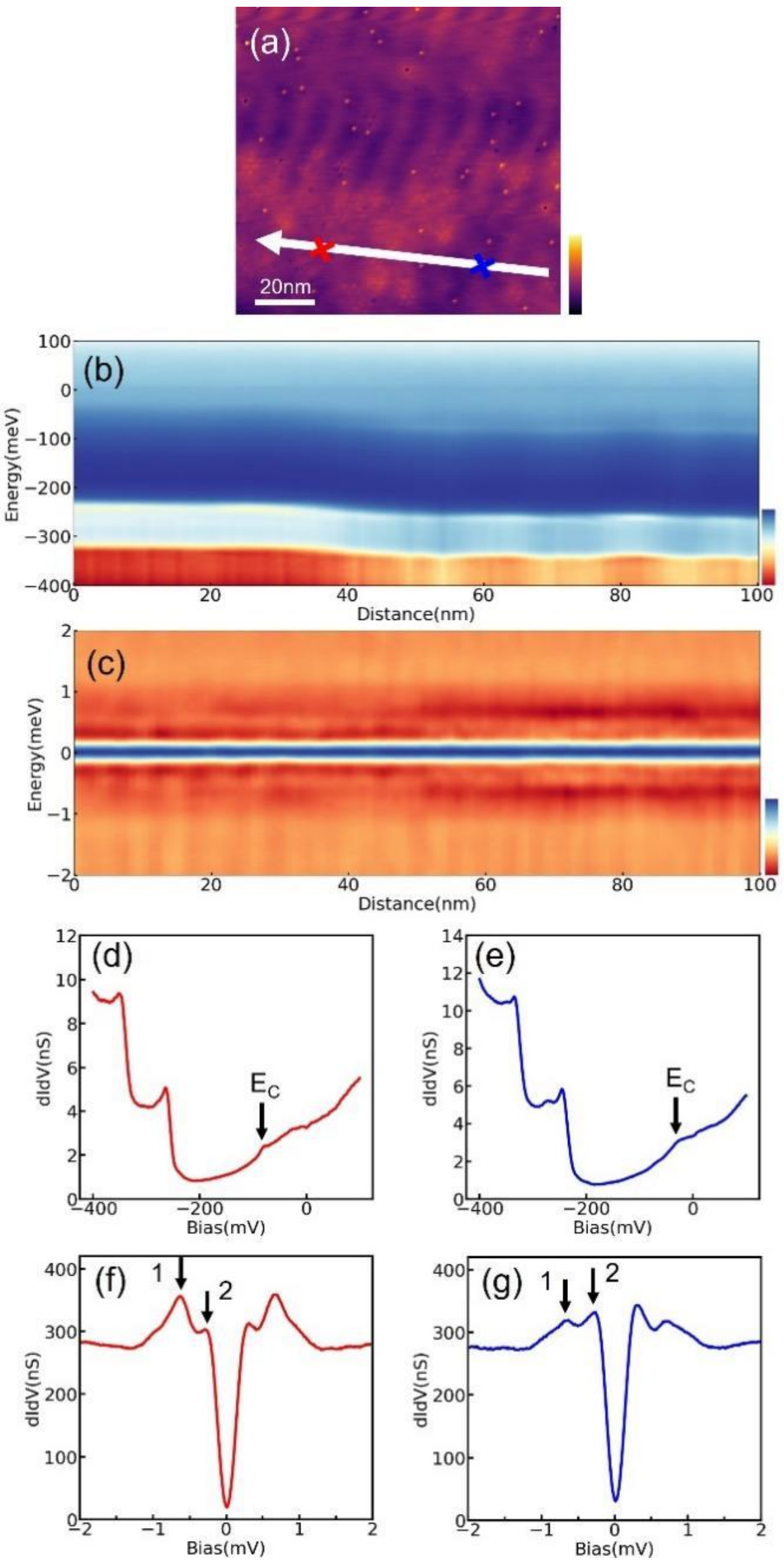

FIG. 3. (a) STM image of the 7QL $Bi_2Te_3$ surface. (b) and (c) spatially resolved dI/dV(x,V) diagrams acquired over a wide and narrow energy range along the white arrows in (a). (d) and (e) Representative dI/dV spectra taken on the two different positions from the line cut in (b). $E_C$ is 80meV and 29meV below the Fermi Level on the two positions respectively. (f) and (g) Representative superconducting spectra extracted from the line cut in (c). The measurement locations are indicated by colored crosses in (a). 1 and 2 indicate the coherent peak and inner peak.

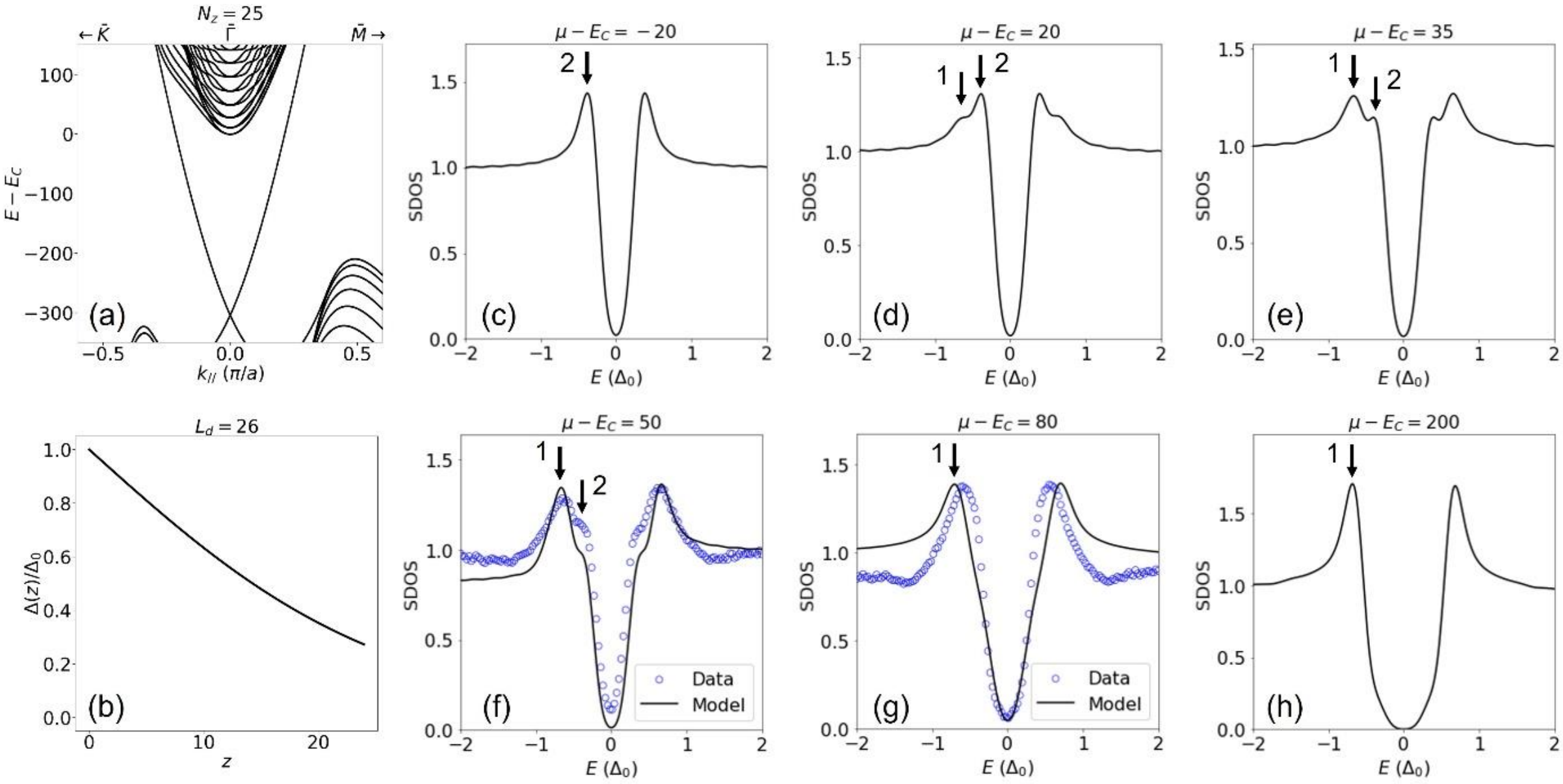

FIG. 4. (a) Simulated band structure of the 5QL $Bi_2Te_3$ thin film. (b) Decay of the proximity-induced superconducting gap that varies as a function of $Bi_2Te_3$ thickness (z). Only a single gap, *i.e.* $\Delta_0$ is considerd. (c) to (h) Surface-projected density of states (SDOS) under various chemical potential $\mu - E_C = -20$, 20, 35, 50, 80 and 200. The blue dots in (f) and (g) are the data for 5QL $Bi_2Te_3$ in Figs. 2(g) and 2(f) respectively

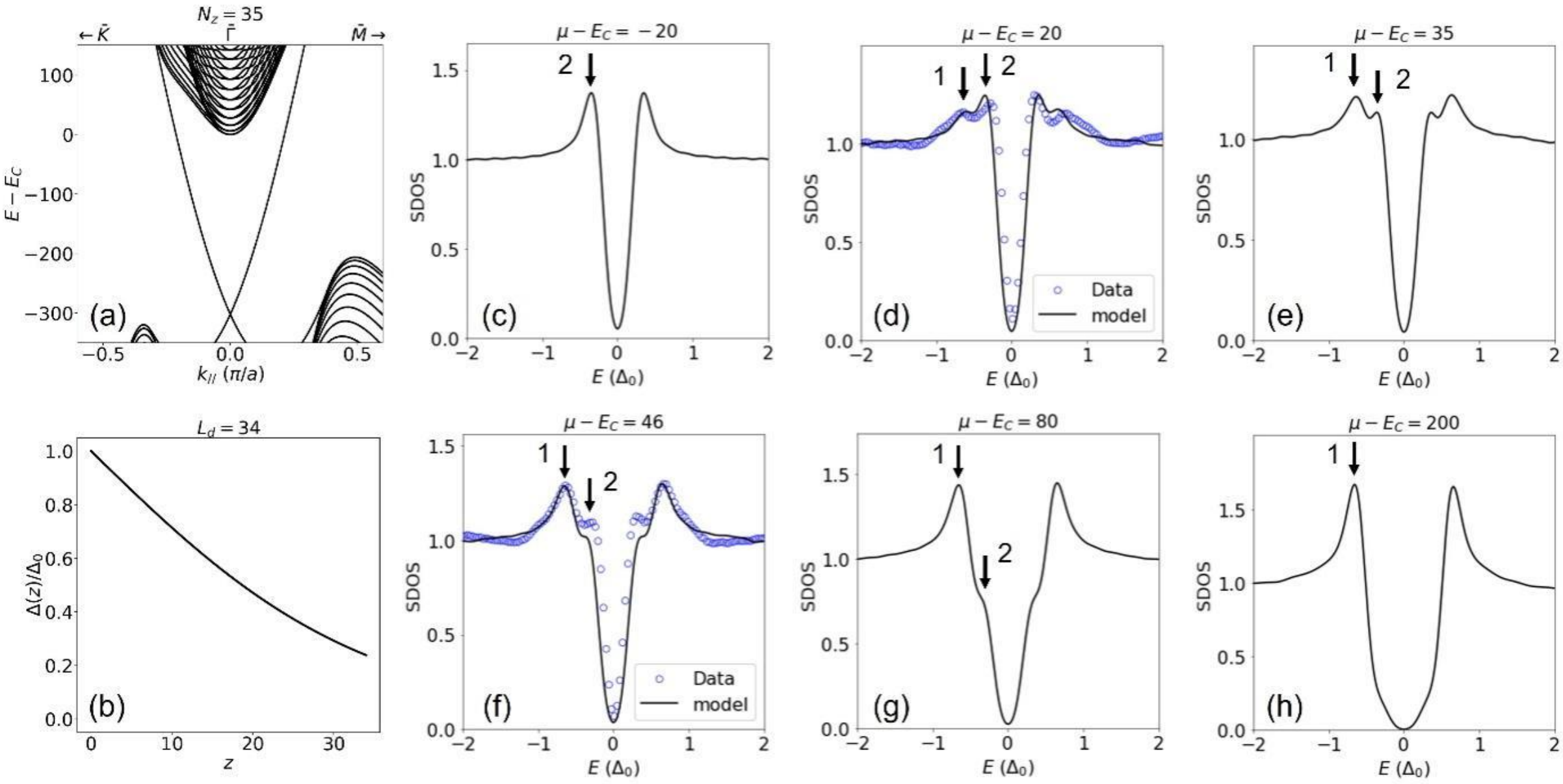

FIG.5. (a) Band structure of the 7QL TI thin film. (b) Proximity-induced superconducting gap that varies as a function of thickness (z). (c) to (h) Surface density of states for various chemical potential $\mu - E_C = -20$, 20, 35, 46, 80 and 200. The blue dots are the data for 7QL $Bi_2Te_3$. The blue dots in (d) and (f) are the data in Figs. 3(g) and 3(f) respectively

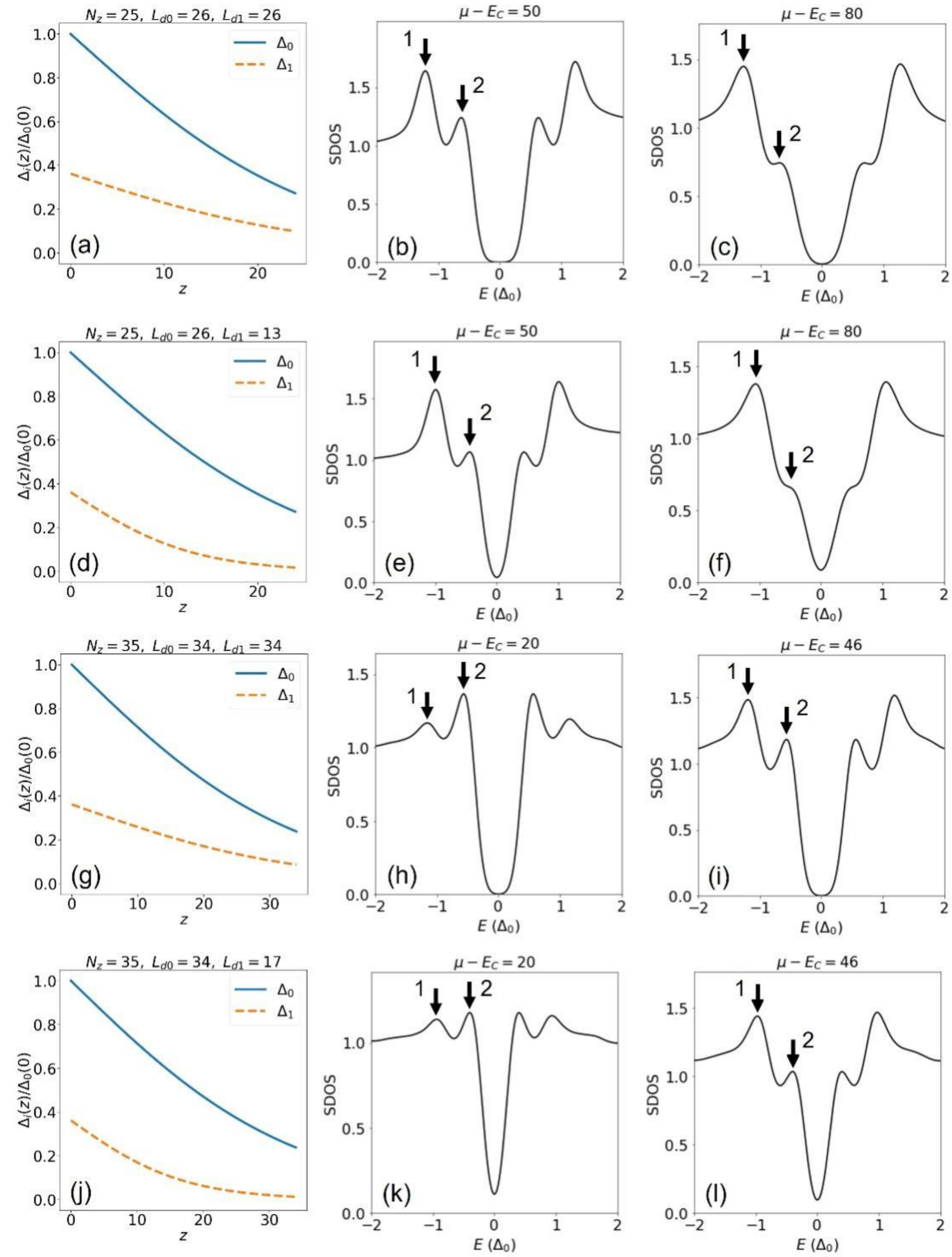

FIG.6. (a) Proximity-induced superconducting gap that varies as a function of thickness (z) for $L_{d0} = 26$ and $L_{d1} = 26$ on a 5QL TI thin film. (b) and (c) The corresponding surface density of states for $\mu - E_C = 50, 80$. (d) Same as (a) but with $L_{d0} = 26$ and $L_{d1} = 13$. (e) and (f) The corresponding surface density of states for $\mu - E_C = 50, 80$. (g) Proximity-induced superconducting gap that varies as a function of thickness (z) for $L_{d0} = 34$ and $L_{d1} = 34$ on a 7QL TI thin film. (h) and (i) The corresponding surface density of states for $\mu - E_C = 20, 46$. (j) Same as (g) but with $L_{d0} = 34$ and $L_{d1} = 17$. (k) and (l) The corresponding surface density of states for $\mu - E_C = 20, 46$.